\documentclass{article}
\usepackage[utf8]{inputenc}
\usepackage{verbatim}
\usepackage{listings}
\usepackage[a4paper, margin=0.7in]{geometry}
\usepackage{braket}
\usepackage{enumitem, kantlipsum}
\usepackage{graphicx}
\usepackage[dvipsnames,table,xcdraw]{xcolor}
\usepackage{multicol}
\usepackage{mathtools}

\usepackage{csquotes}
\usepackage{amsmath}

\usepackage{url}

\usepackage{subfiles}
\usepackage{qcircuit}
\usepackage{amsthm,amssymb}
\usepackage{fdsymbol}
\usepackage{rotating}
\usepackage{caption}
\usepackage{hyperref}

\usepackage{xcolor}
\usepackage{sectsty}
\sectionfont{\color{purple}}
\subsectionfont{\color{teal}}
\subsubsectionfont{\color{olive}}

\usepackage{arydshln}

\usepackage[ruled,vlined]{algorithm2e}

\definecolor{codegreen}{rgb}{0,0.6,0}
\definecolor{codegray}{rgb}{0.5,0.5,0.5}
\definecolor{codepurple}{rgb}{0.58,0,0.82}
\definecolor{backcolour}{rgb}{0.95,0.95,0.92}
\lstdefinestyle{mystyle}{
    backgroundcolor=\color{backcolour},   
    commentstyle=\color{codegreen},
    keywordstyle=\color{magenta},
    numberstyle=\tiny\color{codegray},
    stringstyle=\color{codepurple},
    basicstyle=\ttfamily\footnotesize,
    breakatwhitespace=false,         
    breaklines=true,                 
    captionpos=b,                    
    keepspaces=true,                 
    numbers=left,                    
    numbersep=5pt,                  
    showspaces=false,                
    showstringspaces=false,
    showtabs=false,                  
    tabsize=2
}
\let\OLDthebibliography\thebibliography
\renewcommand\thebibliography[1]{
  \OLDthebibliography{#1}
  \setlength{\parskip}{0pt}
}

\title{\LARGE QKSA: Quantum Knowledge Seeking Agent\\
\large resource-optimized reinforcement learning using quantum process tomography}

\usepackage[affil-it]{authblk}

\author[1,3]{Aritra Sarkar}
\author[1]{Zaid Al-Ars}
\author[1,2]{Harshitta Gandhi}
\author[3]{Koen Bertels}
\affil[1]{Department of Quantum \& Computer Engineering,
\protect\\Faculty of Electrical Engineering, Mathematics and Computer Science,
\protect\\Delft University of Technology, Delft, The Netherlands}
\affil[2]{Guru Gobind Singh Indraprastha University, Dwarka, India}
\affil[3]{QBee.eu, Leuven, Belgium}
\date{} 

\begin{document}

\maketitle

\begin{abstract}

In this research, we extend the universal reinforcement learning (URL) agent models of artificial general intelligence to quantum environments.
The utility function of a classical exploratory stochastic Knowledge Seeking Agent, KL-KSA, is generalized to distance measures from quantum information theory on density matrices.
Quantum process tomography (QPT) algorithms form the tractable subset of programs for modeling environmental dynamics.
The optimal QPT policy is selected based on a mutable cost function based on algorithmic complexity as well as computational resource complexity.
Instead of Turing machines, we estimate the cost metrics on a high-level language to allow realistic experimentation.
The entire agent design is encapsulated in a self-replicating quine which mutates the cost function based on the predictive value of the optimal policy choosing scheme. 
Thus, multiple agents with pareto-optimal QPT policies evolve using genetic programming, mimicking the development of physical theories each with different resource trade-offs.
This formal framework is termed Quantum Knowledge Seeking Agent (QKSA).
A proof-of-concept is implemented and available as an open-sourced software.

Despite its importance, few quantum reinforcement learning models exist in contrast to the current thrust in quantum machine learning.
QKSA is the first proposal for a framework that resembles the classical URL models.
Similar to how AIXI-tl is a resource-bounded active version of Solomonoff universal induction, QKSA is a resource-bounded participatory observer framework to the recently proposed algorithmic information-based reconstruction of quantum mechanics.
Besides its theoretical impact, QKSA can be applied for simulating and studying aspects of quantum information theory like control automation, multiple observers, course-graining, distance measures, resource complexity trade-offs, etc.
Specifically, we demonstrate that it can be used to accelerate quantum variational algorithms which include tomographic reconstruction as its integral subroutine.

\end{abstract}













\section{Introduction} \label{s1}

The theoretical framework of intelligence helps us to understand the capabilities and limitations of natural and artificial intelligence.
Computational learning is being increasingly realized in diverse disciplines.
This fascinating growth 
can however only be sustained by achieving the following three crucial characteristics:
(i) scalability - of computational resources allows applying the system to complex situations;
(ii) explainability - focuses on human understanding of the decision from the learned solution; and,
(iii) generality - involves using a single framework to address multiple scenarios.
Despite the immense success of machine learning approaches, data-driven black-box models currently struggle to address these three aspects in tandem.
In this research, we define a framework that addresses the requirements of these aspects simultaneously.

The holy grail of the field of automation is artificial general intelligence (AGI).
While this was the eventual goal of even the founders of artificial intelligence, AGI has continually alluded computer scientists as a moving target.
Encouraged by the recent achievements of intelligent systems, research on AGI is being revived and pursued from various directions~\cite{domingos2015master}, like, evolutionary approaches, neural networks, and, symbolic logic.
The most theoretically advanced among these is universal artificial general intelligence (UAGI)~\cite{hutter2004universal}.
It is a descriptive theory that is useful for studying super-intelligent AI without building one.
The agent-environment paradigm of model-based reinforcement learning (RL), is best suited to mimic the interactive learning behavior of artificial and biological intelligence.
Lately, the surge of interest in adaptive and autonomous devices has increased the prominence of RL methods beyond robotics and AI communities.
UAGI based RL agents are concisely referred to as universal reinforcement learning (URL) agents.
In this research, we examine policies of modeling an unknown environment as the general task assigned to a URL agent.

URL agents have been instrumental in proving asymptotic optimal behavior in partially observable environments by merging theoretical concepts of decision theory, the notion of a universal automata and algorithmic information theory (AIT).
However, the dependence on AIT makes these agents generally uncomputable.
While resource-bounded variants have been proposed, these models still remain intractable for real world applications.
Moreover, the resource bounds introduce arbitrary hyper-parameters.
To address this issue, we propose to use the idea of embedding RL agents within an evolutionary framework, called EVO-RL~\cite{hallawa2021evo} to guide the hyper-parameter tuning for a specific application scenario.
In this work, we propose for the first time the idea of a resource-bounded EVO-URL.
This is prompted by the suggestion of a UAGI system to eventually play the role of an autonomous scientist by recursive self-improvement (RSI)~\cite{schmidhuber1995learning}.
The RSI characteristics are ensured by embedding the agent's code within a mutating quine.

In defining AGI, the choice of a general environment is as crucial as that of a general learning strategy.
Learning about a physical system by information exchange in its most general form should include classical, quantum, and relativistic scenarios.
In this work, we address the first two cases by defining the environment as an unknown quantum process.
The proposed agent uses quantum process tomography (QPT) as the general algorithm to learn and model the environment.

The major limitation of UAGI is the exponential scaling of the space of programs, which limits its applicability to very simple cases.
To circumvent this, the agent policies in QKSA are chosen from a predefined pool of QPT strategies.
This makes the agent policy computationally tractable as well as explainable, allowing a prescriptive theory of UAGI.
The scalability is bounded by the exponential classical simulation overhead.
This computation cost can be considerably frugal for pragmatic approximation thresholds of classical shadows~\cite{levy2021classical,kunjummen2021shadow} of quantum information.
The proposed \textit{Quantum Knowledge Seeking Agent (QKSA) is an AGI framework based on resource-bounded EVO-URL}.
It models classical and quantum dynamics by merging ideas from AIT, quantum information, constructor theory, and genetic programming.
Following the artificial life (or, animat) path to intelligence, a population of classical agents undergoes open-ended evolution (OEE) to explore pareto-optimal ways of modeling the perceptions from a quantum environment.

Operational effective theories of quantum mechanics have already been reconstructed based on classical information from measurements.
A recent proposal~\cite{mueller2020law} based on Solomonoff induction with the only assumption of computability from algorithmic information theory has been used to reconstruct predictive strategies for non-relativistic classical and quantum environments.
This work is inspired by the \textit{law without law} idea from digital physics.
Using QKSA, we extend this to allow the agents to develop strategies to choose the input states and measurement basis, based on the \textit{participatory observer} notion.
To complement this free choice of the agent, the learning goal is set to reward the predictive capability of environmental interaction while optimizing algorithmic and computational resources.
QKSA does not assume quantum computational capability for the agent in line with the conventional qualia of human intelligence.

This article is organized as follows.
Section~\ref{s2} presents an overview of the URL formalism and recursive self-improvement.
We also present a brief overview of quantum information, computation and tomography.
To bridge the interdisciplinary gap, these background sections are considerably detailed.
Section~\ref{s3} presents the QKSA model and the features.
In Section~\ref{s4} we present formal policy, the specifics of our implementation of this framework.
Section~\ref{s5} demonstrate a proof-of-concept experimental results for process tomography.
In Section~\ref{s6} we conclude the discussion with suggestive future applications.

\section{Conceptual background} \label{s2}

Quantum artificial intelligence (QAI) is an umbrella term exploring the synergy between these two disciplines.
It broadly entails, either, (i) using principles of quantum information and computation within artificial intelligence models, or, (ii) using artificial intelligence for processing quantum information thereby advancing research in quantum technologies.
In our past work~\cite{sarkar2021estimating}, we have proposed general approaches and applications for the former, while in this research, we will focus on the latter case.

Quantum machine learning (QML) is a data-driven sub-field of QAI with a similar bidirectional synergy.
Owing to the huge success of classical machine learning (ML), QML is growing in popularity in recent years.
On one direction, QML generalizes classical models like neural networks, clustering, regression, optimization, etc., to quantum information.
These quantum algorithms running on quantum computers strive for a benefit in terms of runtime, trainibility, solution quality, or memory space, with respect to a classical ML approach.
On the other direction, ML techniques are employed for optimizing and controlling processes in the development of quantum computers.
These include, control of the quantum system, routing and mapping of qubits, quantum error-correction, etc.

Here we note five specific QML solutions which are closely related to QKSA.
\cite{torlai2018neural} uses restricted Boltzmann machines for learning quantum states and processes.
The classical optimizer in quantum variational approaches has also been implemented as a reinforcement learner~\cite{wauters2020reinforcement,yao2020reinforcement,rivera2021avoiding,carrasquilla2021neural}.
Variational quantum algorithms for reinforcement learning using evolutionary optimization~\cite{chen2021variational} has also recently been proposed.
These three implementations are based on data-driven neural networks.
\cite{krenn2016automated} implements a computer algorithm, Melvin, which finds new experimental implementations for the creation and manipulation of complex quantum states.
However the framework is not general for quantum information, and is currently designed for optimizing experiments in quantum optics.
Projective simulation (PS)~\cite{briegel2012projective,melnikov2017active,dunjko2017advances} is a quantum reinforcement learning model that is the most similar in aim to QKSA.
PS is a bio-inspired RL framework which allows the agent, based on previous experience to project itself onto potential future situations, using a stochastic network of clips, called episodic and compositional memory.
It aims to establish a general framework that connects the embodied agent research with fundamental notions of physics. 
Despite the similarities with these approaches, these are not based on a universal computing model.
They are either data-driven heuristics making them not easily explainable, or applied to a specific context and is hard to generalize.
Thus, they do not meet the definition of an AGI agent in a URL setting that we study in this research.

The proposed model of QKSA studies quantum information and computation via the lens of algorithmic information theory (AIT) using reinforcement learning.
QKSA is a generalization of existing classical URL models and thus allows evaluating the computational trade-offs of agency in the formalism of AIT.
Recently, it was shown that it is possible to formulate any problem in AI as reinforcement learning~\cite{silver2021reward}.
Thus, our approach is inherently general for other QAI learning tasks.
The QKSA framework consists of the classical agent which performs the learning activity, and the quantum environment which defines the learning target.
In this section, we first discuss the concepts used in defining a general agent and thereafter, the concepts used in defining a general environment.

\subsection{Classical agent}

The design of the classical agent is based on three otherwise independent concepts from computer science.
At the core is a generalization of a knowledge seeking agent (KSA).
The hyper-parameters that define the resource constraints of the KSA are encoded as a gene.
A population of agents use genetic programming (GP) to evolve by mutation, thereby tuning these parameters.
The KSA and GP is encapsulated within a self-replicating quine, that allows recursive self-improvement.
The background for these three characteristics of the proposed learning agent: knowledge seeking agents, genetic programming and self-replicating programs is explained in this section.

\subsubsection{Knowledge seeking agents for universal reinforcement learning}

Solomonoff's theory of universal inductive inference forms the theoretical basis of UAGI.
It formalizes the two abductive heuristics that are used in scientific modeling, (i) Occam's razor or the principle of parsimony - i.e, when presented with competing hypotheses about the same prediction, select the model with the fewest assumptions, and, (ii) Epicurus' principle of multiple explanations - i.e. retain all theories that are consistent with the observed data.
In this theory, competing predictions are proportionally weighed by the size of the hypothesis that generates the prediction.
This weight is called the algorithmic probability (AP).
To estimate the hypothesis size (or algorithmic information), the environment being modeled is assumed to be computable by a universal Turing machine (UTM).
The algorithmic complexity (AC) thereby defines the hypothesis size.
While the exact values of AP and AC are uncomputable due to the halting problem, upper bound can be estimated using techniques like block decomposition method (BDM)~\cite{soler2014calculating}.
The invariance theorem allows choosing any universal automata or language for the estimation.
This adds a constant overhead based on the cross-compiler program length between the selected machine and the UTM.
For a more pedagogical review of AIT, we refer the readers to \cite{li2008introduction}.

UAGI is formulated in a general reinforcement learning (GRL) setting where the agent and environment interact in turns. 
At every time step, the agent supplies the environment with an action.
The environment then performs some computation and returns a percept to the agent, and the procedure repeats. 
The environment is modeled as a partially observable Markov decision process (POMDP).
The agent cannot observe the underlying Markovian state directly, but receives (incomplete and noisy) percepts through its sensors, and thereby must learn and make decisions under uncertainty in order to perform well.
The canonical model of UAGI is the AIXI model~\cite{hutter2004universal}.
It is the active generalization of Solomonoff induction using Bellman's optimality equation.

Knowledge Seeking Agents (KSA)~\cite{orseau2014universal} generalize the extrinsic reward function in AIXI to a utility function defined as information gain of the model.
Thus, this collapses the exploration-exploitation trade-off to simply exploration, allowing agents to explore the environment in a principled approach.
The goal of these agents are to entirely explore its world in an optimal way and form a model and gets reward for reducing the entropy (uncertainty) in its model from the 2 components: uncertainty in the agent's beliefs and environmental noise.
A particularly interesting case is the KL-KSA~\cite{orseau2013universal}, which is robust to stochastic noise (inherent in quantum dynamics) as the utility function is given as the Kullback-Leibler divergence or relative entropy.

Besides, since UAGI models are only asymptotically computable, it is not a pragmatic algorithmic solution to general RL, and must be simplified in any implementation.
In principle, there are an infinite number of programs that can be candidate models of the environment.
Also, while evaluating, the programs can enter infinite loops.
Thus, to circumvent these two issues, a modified (time and length bounded AIXI) agent called AIXI-tl~\cite{hutter2007universal} limits the length of the programs considered for modeling as well as assigns a timeout for computing the action.
These resource considerations can similarly be translated to the KL-KSA case.

\subsubsection{Genetic programming for resource optimization}

Complementary to the UAGI approach, evolutionary computation uses a different strategy based on biologically inspired models that evolves from a set of simple rules.
It employs a population-based trial and error problem solving technique for meta-heuristic or stochastic optimization.
An initial set of candidate solutions is generated and iteratively updated. 
Each new generation is produced by selecting more desired solutions based on a fitness function, and introducing small random mutations.
The population mimics the behavior or natural selection and gradually evolves to increase in fitness.
Different variants like evolutionary strategies, genetic algorithms, evolutionary programming and genetic programming, were developed to suit specific families of problems and data structures.
There are other metaheuristic optimization algorithms that are also categorized as evolutionary computation, like agent-based modeling, artificial life, neuro-evolution, swarm intelligence, memetic algorithms, etc.

Genetic programming (GP)~\cite{koza1992genetic} is a heuristic search technique of evolving programs, starting from a population of (usually) random programs, for a particular task.
Computer programs in GP are traditionally represented in memory as tree structures (as used in functional programming languages) which can be easily evaluated in a recursive manner. 
The fittest programs are selected for reproduction (crossover) and mutation according to a predefined fitness measure. 
Crossover involves swapping random parts of selected pairs (parents) to produce new and different offspring while mutation involves substitution of some random part of a program with some other random part of a program. 
GP has been successfully used for automatic programming, hyper-parameter optimization, machine learning and in automatic problem-solving engines.
It is especially useful in domains where the exact form of the solution is not known in advance or an approximate solution is acceptable.

\subsubsection{Quines for recursive self-improvement}

The third characteristic that is crucial for the QKSA model is of artificial life (alife).
Alife examines systems related to natural life, its processes, and its evolution, through the use of simulations with (soft) computer algorithms, (hard) robotics, and (wet) biochemistry models.
The field of soft-alife was mainly developed using cellular automata~\cite{neumann1966theory}, while neuro-evolution is another popular technique in use today.
An idea foundational to alife is of a universal constructor.
Universal constructor is a self-replicating automata developed to study abstract machines which are complex enough such that they could grow or evolve like biological organisms.
The simplest such machine, when executed, should at least replicate itself. 
Additionally, the machine might have an overall operating system and extra functions as payloads.
The payload can be very complex like a learning agent or an instance of an evolving neural network.
The design of a self-replicating machine consists of three parts: (i) a program or description of itself; (ii) a universal constructor mechanism that can read any description and construct the machine or description encoded in that description; and, (iii) a universal copier machine that can make copies of any description (if this allows mutating the description it is possible to evolve to a higher complexity).
The constructor mechanism has two steps: first the universal constructor is used to construct a new machine encoded in the description (thereby interpreting the description as program), then the universal copier is used to create a copy of that description in the new machine (thereby interpreting the description as data).

A quine is a program which takes no input and produces a copy of its own source code (and optionally other useful results) as its output.
Thus, it is akin to the software embodiment of constructors.
In principle, any program can be written as a quine, where it (a) replicates it source code, (b) executes an orthogonal payload which serves the same purpose the original non-quine version.

In summary, the crucial elements that will be used for defining QKSA are as follows.
Firstly, the KSA types of URL are used.
These reinforcement learning agents model the environment as programs on a universal automata.
The programs output predictions of subsequent environmental percept when provided with the sequence of past actions and percepts.
The current action is chosen based on the program which has the highest weight determined by (i) having a minimal length (and optionally, by other computational resources constraints like run-time) and (ii) having a high total expected information gain over the time horizon by the sequence of the chosen optimal actions and corresponding predicted perceptions.
Secondly, GP is used for hyper-parameter tuning.
The resource constraints are free hyper-parameters that evolve using mutation between generations of QKSAs.
Thirdly, this EVO-URL framework is encapsulated within the payload of a quine.
This allows the QKSA to be a recursive self-improving agent.
In Section~\ref{s4} we will present the crucial interplay between KSA, GP and Quine for the QKSA framework in more details.



\subsection{Quantum environment}

The definition of a general environment and learning strategy is crucial for AGI research.
In this section we present a brief overview of quantum information and computation, which generalizes classical and probabilistic information processing.
We present quantum process and tomography as the corresponding general environment and learning strategy that will be used by QKSA.

\subsubsection{Quantum computation}

Quantum computation utilizes the laws of quantum mechanics to a computational resource or quality advantage. 
It is the only realizable model of computation that violates the extended Church-Turing thesis, as classical model (like Turing machines) require worst-case exponential resource of space or time to simulate quantum computation.
Information processing via quantum postulates~\cite{nielsen2002quantum} can be described by:
\begin{enumerate}[nolistsep,noitemsep]
    \item 
    The superposition principle defines the possible states $\ket{\psi}$ of a quantum system.
    An isolated system is represented as a linear combination of a chosen orthonormal basis states $\{\ket{i}\}$ with complex coefficients $\alpha_i \in \mathbb{C}$, as $\ket{\psi} = \sum_{i = 0}^{d^n -1} \alpha_i \ket{i}$.
    The complex amplitudes are normalized $\sum |\alpha_i|^2 = 1$, such that $\ket{\psi}$ is a unit vector in an $n$-dimensional complex vector space or Hilbert space. 
    \item 
    The evolution of a closed quantum system is described by an operator $U$ typically associated with the Hamiltonian, i.e. $\ket{\psi'} = U\ket{\psi}$.
    It governs how the state of a quantum system evolves with time. 
    The operator is unitary and thus reversible, $U^\dagger U = U U^\dagger = I$.
    \item 
    The measurement principle governs the collapse of the superposition and bounds the amount of accessible information of a quantum state.
    The Born rule states that the superposition evolves irreversibly to a specific basis state $\ket{i}$ with the probability $|\alpha_i|^2$.
    Quantum measurements are described by a collection $\{M_m\}$ of measurement operators, where $m$ refers to the measurement outcomes and the probability that result $m$ occurs is given $Pr(m) = \bra{\psi}M_m^\dagger M_m \ket{\psi}$.
    The state of the system after measurement is $M_m \ket{\psi} / Pr(m)$.
    While measurements in the physical world are qubit interactions (entanglement) between the system and the environment/detector in the joint Hilbert space, it is yet not possible~\cite{adami2020origin} to derive this postulate from the others.
    \item 
    The state space of a composite physical system $\ket{\psi}$ is the Kronecker tensor product of the state spaces of the $s$-component physical systems, i.e.~$\ket{\psi} = \ket{\psi_1} \otimes \ket{\psi_2} \otimes \dots \otimes \ket{\psi_s}$. 
    Thus the number of parameters needed to describe the state grows exponentially with the number of qubits.
    This is the primary resource in quantum computation to achieve superior computational capability over classical systems by selectively interfering these states by the algorithm.
\end{enumerate}
The logical abstraction of the basis states are typically associated to have a physical meaning.
For example, the simplest quantum states are described as a two-level system, i.e.~$d=2$ called qubits and can be represented by the spin-up/down of an electron.

\subsubsection{Density matrices}

While pure states can be written in terms of ket vectors $\ket{\psi}$, it cannot represent mixed state, i.e.~a statistical ensemble over a set $\{\ket{\psi}_k\}$ of $N$ pure quantum states.
Such states are described as a density matrix $\rho$, as the sum of the probabilities $0 < p_k \le 1$ and $\sum_{k=1}^N p_k = 1$, multiplied by the corresponding projection operators onto certain basis states. 
It is defined as $\rho = \sum_{k=1}^{N} p_k \ket{\psi}\bra{\psi}$.
The bra vector is the adjoint (complex conjugate transpose) of the ket vector, i.e. $\bra{\psi} = \ket{\psi}^\dagger$.
Though the global phase of a quantum state is undetectable, i.e. $\ket{\psi} = e^{i\theta} \ket{\psi}$, a density matrix is unique, as the corresponding global phases of the bra and ket cancel out $\rho = \ket{\psi}\bra{\psi}$.
The corresponding evolution postulate by a unitary transformation $U$ is $\rho' = U\rho U^\dagger$.
A projective measurement of an observable $M_m$ is given by the expectation value $Pr(m) = Tr(M_m \rho)$.
The density matrix formalism deals with observable probabilities whereas ket states deal with complex probability amplitudes.
Statistics of quantum measurements can only estimate the density matrix instead of the state.
Thus, we would use this within the QKSA formalism.

\subsubsection{Quantum processes}

A quantum process $\mathcal{E}$ that transforms a density matrix need not always be unitary.
Given classical processes are often irreversible and include measurements, a quantum generalization includes unitary transforms (symmetry transformations of isolated systems), probabilistic logic as well as measurements and transient interactions with an environment.
Thus, quantum processes formalize the time evolution of open quantum systems.
These are quantum dynamical maps, which are linear and completely positive (CT) map from the set of density matrices to itself.
Typically they are non-trace-increasing maps, and trace-preserving (TP) for quantum channels.
For a quantum system with an input state $\rho_{in}$ of dimension $n\times n$ and an output state $\rho_{out} = \mathcal{E}(\rho_{in})$ of dimension $m\times m$, we can view this system $\mathcal{E}$ as a linear superoperator mapping between the space of Hermitian matrices $\mathcal{E} : \mathcal{M}_{n\times n} \rightarrow \mathcal{M}_{m\times m}$.
While $\rho$ is an order 2 tensor (i.e.~operator), acting on Hilbert spaces of dimension $D=2^n$, $\mathcal{E}$ is an order 4 tensor specified by $D^4-D^2$ parameters.
Beside the superoperator, there are other equivalent~\cite{de2019fault} representations of quantum processes like Choi-matrix $\Lambda$, Kraus operators, Stinespring, Pauli basis Chi-matrix $\chi$, Pauli Transfer Matrix, Lindbladian, etc.

For instance, the Choi matrix $\rho_{Choi}$ is the density matrix obtained after putting half of the maximally entangled state $\ket{\Omega}$ through the channel $\mathcal{E}$, while doing nothing on the other half.
$$\Lambda = \sum_{i,j} \dfrac{1}{2^n} \ket{i}\bra{j} \otimes \mathcal{E} (\ket{i}\bra{j})$$
$$\rho_{Choi} = \Lambda (\ket{\Omega}\bra{\Omega})$$
Thus it requires $2n$ number of qubits, but since the input state is fixed, we effectively do a quantum state tomography (QST) on this larger space instead of QPT, reducing the overall number of trials.
The evolution of a density matrix with respect to the Choi-matrix is given by:
$$\rho_{out} = \mathcal{E}(\rho_{in}) = \text{Tr}_1((\rho_{in}^T \otimes \mathrm{I}) \rho_{choi}))$$
where $\text{Tr}_1$ is the partial trace over subsystem 1.
As a result of the Choi-Jamiolkowski isomorphism, the Choi matrix $\Lambda $ characterizes the process $\mathcal{E}$ completely.
This forms the basis of the channel-state duality between the space of CP
maps and the space of density operators. 

\subsubsection{Quantum tomography}


Characterization of quantum dynamical systems is a fundamental problem in quantum information science. 
Several procedures that have been developed that achieve this goal are called quantum process tomography (QPT).
Some examples of QPT techniques are: standard quantum process tomography~\cite{chuang1997prescription}, entanglement-assisted quantum process tomography (EAQPT), direct characterization of quantum dynamics, compressed-sensing quantum process tomography, permutation-invariant tomography, and self-guided quantum process tomography~\cite{hou2020experimental}
Each QPT technique have different experimental setup and computational resource requirements.
An exhaustive survey of all techniques considering an inclusive figure-of-merit with respect to resources and their tradeoffs have not been undertaken previously.
\cite{mohseni2008quantum} provides a good overview of some of the most used techniques and comparison in terms of the resources of the Hilbert space, input state complexity, required measurement and required interactions.
In the proof-of-concept implementation of QKSA, we will use EAQPT for the experimental results.
EAQPT is based on the Choi-Jamiolkowski isomorphism, as it uses QST to reconstruct the Choi density matrix of the quantum process.


A more recent development towards the limits of quantum tomography is based on classical shadow of states~\cite{aaronson2019shadow,huang2020predicting} and processes~\cite{levy2021classical,kunjummen2021shadow}.
Shadow tomography aims to extract key information about a state/process with only polynomially many measurements.
These form a good set of candidate QPT algorithms that can be used to explore the resource trade-offs for limiting cases.

As a summary, within the QKSA formalism, QPT reconstruction algorithms form the space of programs that are evaluated by the agent as candidates for the modeling of the environment.
Given computational resource limitations, QKSA can automatically discover the optimal strategy in the available pool of QPT algorithms.
In canonical UAGI formalism, the pool of programs are drawn randomly from a prefix-code for an universal automata.
However, the space of programs grows exponentially, limiting its applicability.
We restrict this space to a constant number of predefined algorithms.
Intuitively, a QPT algorithm will perform better in predicting a quantum environment than a random program.
Thus, it allows us to apply the tools of AIT in practical setting where available expert knowledge can be embedded within the agent, making it tractable.


\section{Characteristic features of QKSA} \label{s3}

\subsection{Classical observers in a quantum world}


Similar to UAGI, in digital physics~\cite{zuse1970calculating}, the universe is modeled as a vast, (quantum) computation device, or as the output of a deterministic or probabilistic computer program.
Information is increasingly put into the central stage in physics, especially in reconstructing theories like quantum mechanics from general principles~\cite{hardy2001quantum, masanes2011derivation,hohn2017quantum} as well as it's physical nature~\cite{vopson2019mass}.
John Archibald Wheeler~\cite{wheeler2018information} popularized this idea as \textit{it from bit}.
This meant that every item of the physical world has at bottom an immaterial source and explanations of what we call reality arises from the posing of yes-no questions and the registering of equipment-evoked responses.
Quantum information theory as a generalization of Boolean logic is used by Seth Lloyd~\cite{lloyd2006programming} to extend this principle, with the evolution of the universe as an ongoing quantum computation, with the fundamental laws of physics constituting the program.
Wheeler asked the question of the possibility of the existence of an ultimate law of physics, from which everything that is knowable about the material world can be deduced.
This idea has been coined as law without law~\cite{deutsch2017wheeler}. 
If such a principle does not exist, it would mean that certain aspects of the natural world are fundamentally inaccessible to science.
Instead, the existence of such a unified law would have to explain its own origin and preferential bias.
So, paradoxically, the ultimate principle of physics, cannot be a ``law" (of physics), hence the expression.
Thus epistemological assumptions of how physical theories are formed and verified becomes imperative, removing physics as the science of ``what is" to that of ``what we observe".

The recent work from Markus Müller~\cite{mueller2020law} is central to the ideas developed in this research.
It claims that given a complete description of the current observer state $x$, it is possible to predict what state $y$ the observer will subsequently evolve to using $P(y|x)$ based on Solomonoff's algorithmic probability, universal prior and universal induction.
This currently encompasses classical (non-relativistic) and quantum physics, and can be used to reconstruct an operational theory based on this assumption of the world being computable, and there are no super-Turing physical processes.
Note this is entirely an epistemic derivation of the laws of physics based on information axioms, and focuses on using the theories to predict the results of future interactions, instead of assuming any ontological interpretations of the model.

In this research we extend the research from~\cite{mueller2020law} to an implementation of a framework to allow experimentation.
While doing so, we narrow down on the specifics of the original ideas.
The primary objective of AGI models like QKSA is to mimic human behavior to form explainable hypothesis about the environment.
Semantic explanation based on human qualia is represented in terms of classical information.
This does not restrict representing quantum information, as using the standard formalism we can represent quantum information using a worst-case exponential amount of real-valued classical information.
Thus, the QKSA specifically models classical observers in a quantum world, and still recover and learn features that can help us form hypothesis and predict the environmental dynamics.

\subsection{Quantum process tomography as a general modeling technique}

Extending Müller's idea to the Church-Turing-Deutsch thesis, the program (that the Solomonoff induction uses) is basically an efficient quantum computing simulator, given a classical computing substrate.
This can also be a programmable quantum simulator~\cite{feynman2018simulating} given a quantum computing substrate.
A model of the environment (universe) is created from the agent's (observer's) perspective, representative of the black-box input-output behavior of the environment.
Given knowledge of the environmental dynamics, it is possible to create the corresponding classical model (e.g. a Grover search simulator). 
However, for unknown environment, the general technique is to do process tomography, thus, that is the general modeling algorithm we would focus on.
For the rest of this article, we consider the general case of quantum environments.
A classical environment can be efficiently mapped to a corresponding quantum environment.

For the general quantum case, what kind of algorithms would execute for predicting the next observer state using Solomonoff universal induction?
Given that it is possible to simulate quantum physics on a classical simulator (albeit by incurring exponential resource cost), the good predictor will be a quantum process tomographic reconstruction based on the previous observer states. 
Thus, an agent trying to derive the information based operational laws of quantum physics would converge to a QPT algorithm as the best predictor for subsequent environmental percepts.
And thus, to define a tractable formalism, we can consider the subset of all QPT algorithms instead of the entire space of programs for the universal automata.

\subsection{Participatory observer as a reinforcement learning agent}

Consider the phase before the process matrix has been reconstructed up to a certain degree of precision (i.e. before an informationally complete history of observations is registered). 
In this phase, the participatory observer can choose an action like a UAGI agent based on the process matrix.
The subsequent perception will be based on both the chosen action and the environment. 
Thus, this phase is not fully modeled by Solomonoff's induction. 
In the reinforcement learning setting, the next observer state is based on both the current observer state (that defines the memory of previous observations and the current action based on the QPT scheme) as well as the part of the environmental dynamics that has not yet being learned. 

Given a complete description of the environmental dynamics already learned and encoded within the observer state $x$, it is still not possible for the agent to perfectly predict individual measurements.
However, with access to an additional random tape, the statistical distribution of measurement results in any chosen basis can be predicted by the agent.
Thus instead of predicting individual perceptions, a quantum UAGI agent can only predict the expected probability of a measurement.
This is inherently dependent on the choice of a measurement basis from the agent. 
This is called the participatory observer principle which states - physical things are information-theoretic in origin and this is a participatory universe where the interactions define the reality we perceive.
QKSA in effect learns efficient strategies for observer participancy for modeling quantum environment.

\subsection{Open-ended evolution of pareto-optimal computational resources}

The algorithmic probability of the program is used as weight for the chosen action and the reward in UAGI.
However, this also makes such models impractical due to the uncomputability of algorithmic information metrics like algorithmic probability and algorithmic complexity.
Thus, pragmatic implementation of these models like AIXI-tl, MC-AIXI$_{\text{(FAC-CTW)}}$ and UCAI~\cite{katayama2018computable} bounds the program length and runtime per step to explore a subspace of promising hypotheses that models the interactive behavior registered till the current time step.
There arises three issues with this approach:
\begin{enumerate}[nolistsep,noitemsep]
    \item The bounds introduce heuristic hyper-parameters that depend on the available computational resources. Thus it becomes difficult to select an appropriate value to apply the model for a given use-case.
    \item The bounds sharply cut off models beyond the specification while keep the weight for the models within the specification unaffected. So a model that performs well but just lies beyond the defined bound may be unreachable.
    \item It is possible to trade off these resource bounds with other computational resources, like additional memory.
\end{enumerate}

Using the QKSA platform, it is possible to investigate these issues.
In the framework we consider five computational resources together called the \emph{LEAST} metric, as an acronym for (program/hypothesis) \underline{l}ength, (compute) \underline{e}nergy, \underline{a}pproximation, (work memory) \underline{s}pace and (run) \underline{t}ime.
Similar algorithmic observables has been suggested in \cite{baez2012algorithmic}.
We provide estimation techniques  of the \emph{least} metric for each based on state-of-the-art algorithmic information research and general practices in computer engineering.
The estimation technique however can easily be redefined by the user.
These estimated metrics are used in a two-fold way.
Firstly, it is used to qualify the hypothesis for consideration based on an upper bounds for each of the five metric individually.
This is dictated by the available computational resource and similar to the resource-bounded UAGI models.
These bounds can be included in the list of evolving hyper-parameters to allow QKSA to mutate and adjust autonomously to the available computational resource.
Thereafter, the metrics for valid hypothesis are fed to a cost function (a genetic program) that outputs a single positive real value which is used as the weight for the hypothesis in the semi-measure instead of only the length, as in algorithmic probability.
We call this the \emph{least action} as a parameterization to optimize the Lagrangian dynamics of within computational space-time.

Currently these is no unifying cost function that can serve as a metric to trade-off bounds on resources (like space, time, approximations).
In fact this depends closely on the policy of the agent.
For examples, a physicist might choose to use simpler Newtonian mechanics instead of complex relativistic mechanics for modeling where the approximations are acceptable.
Thus, instead of a single metric, a pareto-optimal frontier on the least metrics maps to models and algorithms that can be used to predict the environment dynamics.
Various research has explored this frontier considering a few of the least metrics.
For example, Bennett's logical depth and Schmidhuber's speed prior trades off time-length; Wolpert's research deals with the thermodynamic complexity of Turing machines; look-up tables trades off time-space, etc.

QKSA holistically explores these trade-off via the GP function.
The five estimates of the least metrics are input to a cost function.
The cost function itself is a gene represented as a program tree with the leaf nodes as the metrics or constants and the internal nodes are from a set of basic arithmetic functions (addition, multiplication, square-root, logarithm, etc.).
Once QKSA learns an environment optimally or completely fails to learn the environment (i.e. when the learning rate stabilizes), the QKSA self-replicates by invoking the quine functionality.
The child QKSA has the same source code as the parent, except a mutation on the cost function that modifies the weights and structure embedded via the cost function gene.
Thus the open-ended evolution of the pareto-optimal manifold converges on QPT algorithms which fits well in the available computational resource.
The parent QKSA perishes if the prediction of the model fails persistently (i.e. when the rate stabilizes as the strategy fails to learn) or continues to correctly predict environmental interaction and can be inspected to obtain the cost function.

\subsection{Utility function as quantum complexity distance}

The learning process in RL is guided by a reward function assigned by the environment as part of the perception.
While this is trivial to define for game environments, it is difficult to define for modeling environment dynamics without introducing another third-party evaluating agent.
To circumvent this, we consider the generalization of rewards, as utility computed by knowledge seeking agents, instead of being an external input.
The utility is a metric estimated internally by the agent based on a self-defined distance measure in the space of percepts.
As already discussed, due to the inherent randomness of quantum measurements, it is not possible to predict individual measurements in arbitrary basis even with full knowledge of the system.
Thus, the metric is evaluated on the stochastic distribution of percepts.
The process matrix reconstructed by the QPT algorithm from the already known history of action and corresponding percept be called $\rho_{Choi}^t$.
An update of this matrix $\rho'_{Choi}$. is predicted based on the current chosen action $a_t$ and the corresponding prediction of the percept $e'_t$.
The actual update $\rho_{Choi}^{t+1}$ is however based on the actual perception from the environment $e_t$.
The distance measure between these two updates is the utility.
Once the process matrix is fully learned, this distance will converge.
Thus, QKSA is a generalization of KL-KSA for density matrices.


Unlike classical probability distribution, there are many measures of quantum distances each with their own application advantage.
The QKSA framework allows the user to select a distance metric as part of the experimental setup.
The current version provides the following distance metrics, Hamming distance, KL divergence, trace distance, Hilbert-Schmidt norm and Bures distance (fidelity).
Users can also define their own custom distance measure.
Future extension would provide diamond distance, Hellinger distance, quantum Kolomogorov complexity, quantum relative entropy, Rényi divergence, Bhattacharyya distance and quantum complexity action~\cite{halpern2021resource}.


\section{QKSA framework} \label{s4}

In this section the formal framework of the QKSA is presented by defining the parameters within an implementation that captures the agent scheme discussed so far.
Thereafter, an execution procedure and the framework is outlined.

\subsection{Parameters definitions}

The standard formalism of reinforcement learning includes:
\begin{itemize}[noitemsep,nolistsep]
    \item $t_p$ is the number of time steps in the past that is considered by the agent at each point in time. In AIXI, this considers all steps since the inception of the agent. For pragmatic implementations and in QKSA, it is typically a sliding window of few steps in the recent past based on the available total memory of the agent.
    \item $t_f \in \{1,\infty\}$ is the number of time steps that the agent predicts in the future, or, the remaining duration the agent is run. It is also called the horizon. In the limiting case, the number of steps for asymptotic convergence to the optimal strategy is infinity. For QKSA we will consider only 1 step in the future but the implementation is generic and can be extended to any number of steps.
    \item $a_t \in \mathcal{A}$ is the chosen action from the action space at time step $t$.
    \item $e_t \in \mathcal{E}$ is the perception recorded by the agent at time step $t$ from the percept space.
    \item $e'_t \in \mathcal{E}$ is the prediction of the perception $e_t$ made at time step $t-1$.
    \item $\lambda^{e'_t} \in \{0,1\}$ is the probability of the prediction $e'_t$ made at time step $t-1$.
    \item $h_t$ is the sequence of the history of actions and perceptions up to time step $t-1$. It implemented as a ring buffer of $h_t = a_{t-t_p}e_{t-t_p}\dots a_{t-1}e_{t-1}$
    \item $\rho_t$ is the hypothesis or model of the environment generated by processing the history $h_t$ by a candidate QPT reconstruction algorithm. It is typically a Choi matrix of the learned environmental quantum process.
    \item $p_{qpt}$ is the QPT program that is executed on the defined computational model $C$. It is capable of, (i) generating a tomographic reconstruction $\rho_t$ of the environment given the history $h_t$, (ii) provide an expectation value of a prediction $e'_t$ given a $\rho_t$ and $a_t$.
\end{itemize}

The \emph{least} metric defines the bounds on the hypothesis-space and the relative weight assigned to each considered hypothesis.
It takes into account the 5 cost metrics of program length, thermodynamic cost, approximation, space/memory and run-time.
The hypothesis-space of QPT  is bounded by the 5 $\text{least}_{max}$ hyper-parameters.
All trial hypotheses must lie within the bounds of all 5 parameters.
Once a trail hypothesis is admitted based on the $\text{least}_{max}$ bounds, the estimate of the 5 cost parameters $\text{least}_{est}$ is combined to form a single indicative metric of the fitness of the hypothesis.
Each parameter has an associated weight or scaling factor $w_{\text{least}}$.
The cost function defines the equation to combine the $\text{least}_{est}$ and $w_{\text{least}}$, and is subject to evolution.
\begin{itemize}[noitemsep,nolistsep]
    \item $d$ refers to the data on which the least metric is evaluated. It consists of the history $h_t$ and the QPT algorithm $p$. The cross-compiler description length between the chosen automata $C$ (a Python compiler) and the canonical UTM is assumed to be constant and ignored for relative evaluation.
    \item $l_{max}$ is the maximum length of $p$ that is considered.
    \item $e_{max}$ is the maximum energy cost of executing $p$ for the functions discussed above.
    \item $a_{max}$ is the maximum approximation threshold used by $p$ for the functions.
    \item $s_{max}$ is the maximum space or working memory that $p$ can use while execution. It typically includes the $h_t$ as well.
    \item $t_{max}$ is the maximum execution time for $p$ before it generates the output for the functions.
    \item $l_{est}$ is an estimate of the length of $p$ that is considered, that outputs $d$. A rough estimate can be arrived at by the bit length of the QPT program. Lossless compression or BDM~\cite{soler2014calculating} can also be used for a more tight estimate.
    \item $e_{est}$ is an estimate of the energy cost of executing $p$. Research into this aspect is scarce, specially for high-level programs. The recent proposal on the thermodynamic Kolmogorov complexity~\cite{kolchinsky2020thermodynamic} needs to be explored further for estimating the energy cost. It can also be externally estimated by the energy consumption of the computational automata $C$.
    \item $a_{est}$ is an estimate of the deviation arising from approximations made by the program $p$.
    \item $s_{est}$ is an estimate of the space or working memory that $p$ uses while execution.
    \item $t_{est}$ is an estimate of the execution time for $p$ before it generates the output.    
    \item $w=\{w_l,w_e,w_a,w_s,w_t\}$ is a set of associated weights for each of the least metrics.
    \item $c_{least}$ is the cost function that takes in the 5 least metrics and a weight for each metric and calculates a cost based on the evolving gene.
    \item $m_c$ is the mutation rate of the cost function $c_{least}$.
    \item $F$ is the set of functions allowed in the cost function $c_{least}$ and typically includes standard operations like addition, multiplication, exponentiation, logarithm, etc.
    \item $c_{est}$ is the estimated cost based on the estimated least metrics and the cost function. This is the generalization of the program length as used in UAGI.
\end{itemize}

The parameters for the quine define the learning progress and when the agent self-replicates.
Replication is triggered based on the fitness of the hypothesis based on the predictive capacity over time.
\begin{itemize}[noitemsep,nolistsep]
    \item $\Delta$ is a distance measure (e.g. Hamming distance, trace distance) defined between process matrices $\rho$.
    \item $u'_t$ is the predicted utility between the predicted update to the process and the current learned process. It is the relative distance using $u'_t = \Delta(\rho'_{t+1},\rho_t)$.
    \item $\gamma_t$ is the discount that is proportional to the time span $t$ between the predicted reward/utility step and the current time step. It depends on the dynamic and episodic nature of the environment. For time steps further in the future, prediction penalties can be scaled down. For episodic environment, the value is 1 for the next time step and 0 otherwise.
    \item $R_t=\sum_{e'_t\in\mathcal{E}} \lambda_t \dots\sum_{e'_m\in\mathcal{E}} \lambda_m \sum_{k=t}^m \gamma_k u'_k$ is the cumulative discounted return at time step $t$. Note that since the QPT algorithms chooses predictions probabilistically, the weighted summation over the sequence of predictions needs to be considered.
    \item $u_t$ is the actual utility between the interaction steps. It is the relative distance $u_t = \Delta(\rho_{t},\rho_{t-1})$.
    \item $K_t=u'_t - u_t$ knowledge at time step $t$.
    \item $K_R$ is the knowledge threshold for reproduction. If $K_t < K_R$, the agent self-replicates with mutation in its hyper-parameters.
    \item $K_D$ is the knowledge threshold for death. If $K_t < K_D$ the agent halts (dies).
\end{itemize}

\subsection{Agent formalism}

The main advantage of using universal reinforcement learning is that it allows us to define the learning model mathematically.
In this section we start from the formalism of the classical KL-KSA and elucidate the changes that leads to the QKSA formalism.

For simplicity, the action and percept spaces are assumed to be stationary (i.e. time-independent and fixed by the environment) and countable (although most results generalize to continuous spaces).
The agent is formally identified by its policy $\pi$, which in general is a distribution over actions for the current step, conditioned over the history, denoted by, $\pi(a_t|h_t)$.
The environment is modeled as a distribution over percepts, $\nu(e_t|h_t a_t)$.
A rational agent based on Von Neumann-Morgenstern utility theorem strives to maximize the expected return, called the value.
The value achieved by a policy in an environment given a history is defined as: $V_\nu^\pi(h_t)=\mathbb{E}_\nu^\pi\big[R_t\big|h_t\big]$.
This can be expressed recursively, as the Bellman optimality equation, $$V_\nu^\pi(h_t)=\sum_{a_t\in\mathcal{A}}\pi(a_t|h_t) \sum_{e_t\in\mathcal{E}}\nu(e_t|h_t a_t) \big[\gamma_t r_t + \gamma_{t+1} V_\nu^\pi(h_{t+1})\big]$$
AIXI-based models use Solomonoff's universal prior for mixing over the model class $\mathcal{M}$ of all computable probability measures using the Kolmogorov complexity of the environment, $w_\nu = 2^{-l(\nu)}$.
The environment is usually modeled as programs on a UTM, denoted as U, typically a monotone TM with 3 tapes, for input (perception), working and output (action).
Thus, $\xi(e_t|h_t a_t) = \sum_{\nu\in\mathcal{M}} w_\nu \nu(e_t|h_t a_t)$ and the optimal policy maximizes the $\xi$-expected return, i.e. $\pi^{\text{KL-KSA}}=\arg\max_\pi w_\nu V_\xi^\pi$.
Distributing the $\max$ and $\sum$ in the recursive equation yields the canonical expectimax equation as, $$a_t = \arg \lim_{m\rightarrow \infty} \max_{a_t\in \mathcal{A}}\sum_{e_t\in\mathcal{E}}\dots\max_{a_m\in \mathcal{A}}\sum_{e_m\in\mathcal{E}} \sum_{k=t}^m \gamma_k u_k \sum_{q:U(q;a_{<k})=e_{<k*}} 2^{-l(q)}$$
In the case of KL-KSA, the reward for AIXI is generalized to the utility given by $$u(e_t|ae_{<t}a_t) = Ent(w_\nu|ae_{<t+1})-Ent(w_\nu|ae_{<t}a_t)$$

The first change is to restrict the search space of programs $p$ to quantum process tomography algorithms, denoted as $p_{qpt}$.
Strictly there is no need to specialize the search to this subspace of program.
Searching over the full space of programs would lead to higher rewards for QPT algorithms owing to their predictive capability and thereby select actions based on this subspace.
However since we are interested in a pragmatic implementation, searching the full space of programs is intractable even for very modest cases.
It is important that the QPT algorithm reconstructs and outputs a process representation $\rho_t$ instead of the prediction of the next perception.
This is imperative due to the stochastic nature of individual quantum measurement and the calculation of the utility.

The second change is to replace the length estimate of the $2^{-l(p)}$ factor from the algorithmic probability with the estimate of the evolving cost function $c_{est}$.
The cost function is denoted by $c_{least}$, i.e. $c_{est} = c_{least}(p_{qpt})$.
Thus, the learning part of the equation is:

$$\boxed{ a_t^{QKSA} = \arg \lim_{m\rightarrow \infty} \max_{a_t\in \mathcal{A}}\sum_{e'_t\in\mathcal{E}} \lambda^{e'_t} \dots\max_{a_m\in \mathcal{A}}\sum_{e'_m\in\mathcal{E}} \lambda^{e'_m} \sum_{k=t}^m \gamma_k u'_k   \sum_{\substack{p_{qpt}:U(p_{qpt};h_k)=\rho_k\\p_{qpt}:U(p_{qpt};\rho_k;a_k;e'_k) = \lambda^{e'_k}}} 2^{-c_{least}(p_{qpt})} } $$

\begin{figure}[ht]
\centering
\includegraphics[clip, trim=9cm 0cm 0cm 8cm,width=0.8\textwidth]{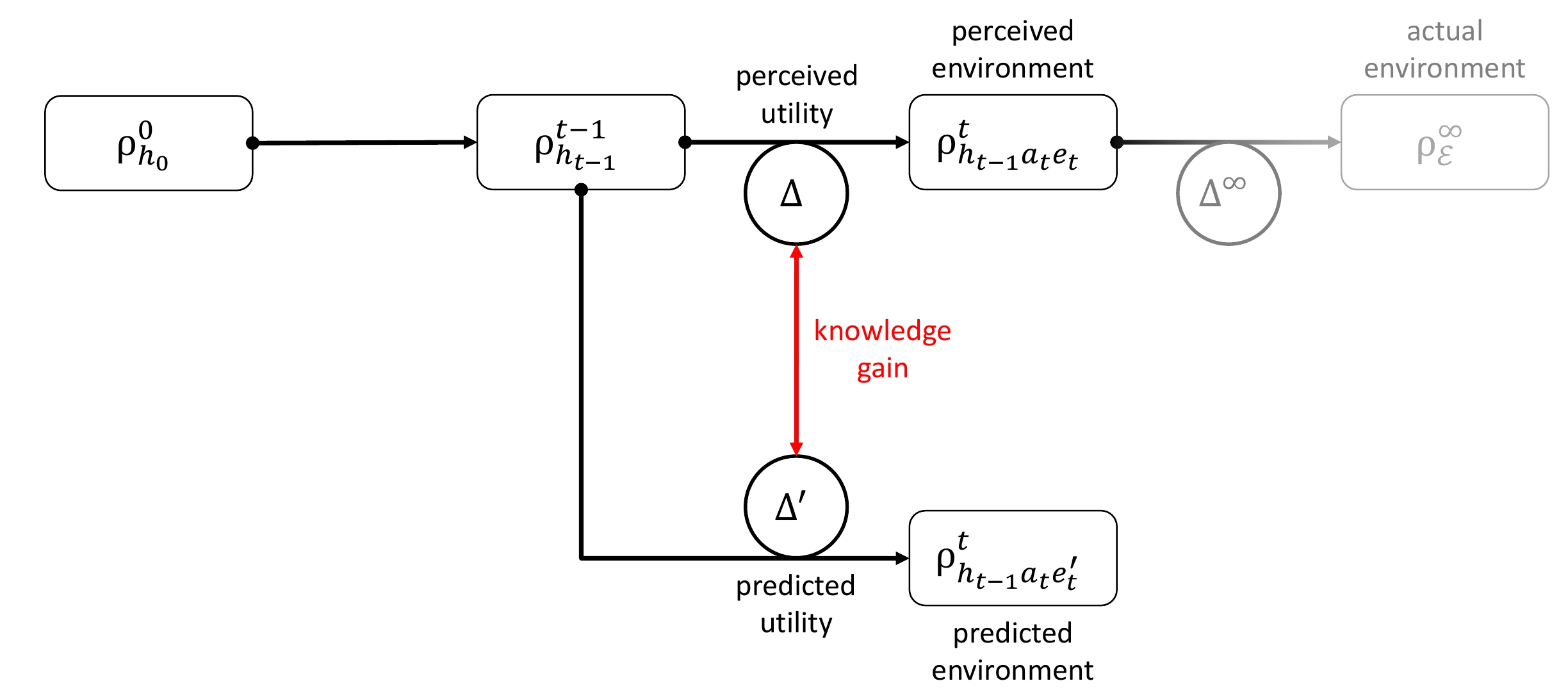}
\caption{QKSA knowledge gain}
\label{fig:kg}
\end{figure}

The third change is to define the utility function as a quantum distance measure on the space of quantum processes $\rho$ defined as the density matrix in the Choi process matrix representation.
A higher predicted utility indicates that the current estimate of the process will be updated more significantly based on the perception, thus, a potential knowledge gain for choosing that action.
These relationships are shown in Figure~\ref{fig:kg}.
$$\boxed{ u'_t = \Delta(\rho'_{t+1},\rho_t) = \Delta(U(p_{qpt};h_t ;a_t; e'_t),U(p_{qpt};h_t)) }$$

\subsection{Episodic environment}

While the infinite horizon is used for proving asymptotic optimality, a finite horizon is required for any pragmatic implementation.
Since the QPT environment is episodic, i.e. the environment is reset after each interaction cycle, a horizon of 1-step captures the highest possible level of temporal dependency.
The simplification of the policy for 1-step horizon, i.e. $k = m = t$ is:
$$\boxed{ a_t^{QKSA} = \arg \max_{a_t\in \mathcal{A}}\sum_{e'_t\in\mathcal{E}}  \lambda^{e'_t} \Delta(\rho'_{t+1},\rho_t) \sum_{\substack{p_{qpt}:U(p_{qpt};h_t)=\rho_t\\p_{qpt}:U(p_{qpt};\rho_t;a_t;e'_t)=\lambda^{e'_t}}} 2^{-c_{least}(p_{qpt})} }$$


Let us understand this simple case in more depth.
At each step, the QKSA algorithm consists of two phases, learning and evolution: 
\begin{itemize}[nolistsep,noitemsep]
    \item In the learning phase:
    \begin{enumerate}[nolistsep,noitemsep]
        \item A pool of QPT algorithms is inspected.
        \item Each QPT algorithm is used to reconstruct the unknown environment as a process matrix based on the history of actions and perceptions.
        \item This reconstruction process is used to estimate the resource cost of the QPT algorithm.
        \item The process matrix is then used on all possible actions the agent can take at this step.
        \item For each such action, the process matrix predicts a distribution of perceptions. Thus, for each action-prediction pair, the process matrix generates a probability for the prediction using the QPT algorithm.
        \item Each predicted perception would lead to a predicted update for the process matrix.
        \item These predicted process matrices are compared with the current process matrix to generate the predicted utility of the corresponding action-prediction pair.
        \item These predicted utilities are weighted by the probability of that specific prediction that led to the predicted utility.
        \item These utilities for a specific action are accumulated as the utility for the action for all possible predictions of perceptions.
        \item This sum of utility for an action is weighed by the resource cost of the QPT algorithm used for modeling and prediction.
        \item The action that maximizes this weighted value is chosen as the action for the current step.
    \end{enumerate}
    \item In the evolutionary phase:
    \begin{enumerate}[nolistsep,noitemsep]
        \item The utility is used to calculate the return over the number of prediction steps based on the weights for each prediction used to calculate the utility.
        \item The total return is the learning gradient.
        \item If this return is below a threshold, the agent reproduces by mutating the cost function and self-replicating.
        \item Alternatively, if the return is above a threshold, the agent dies. There is also a maximum limit on number of interaction steps and number of reproductions after which the agent halts.
    \end{enumerate}
\end{itemize}

\subsection{Execution procedure}

\begin{figure}[h]
\centering
\includegraphics[clip, trim=44cm 0cm 0cm 60cm,width=0.99\textwidth]{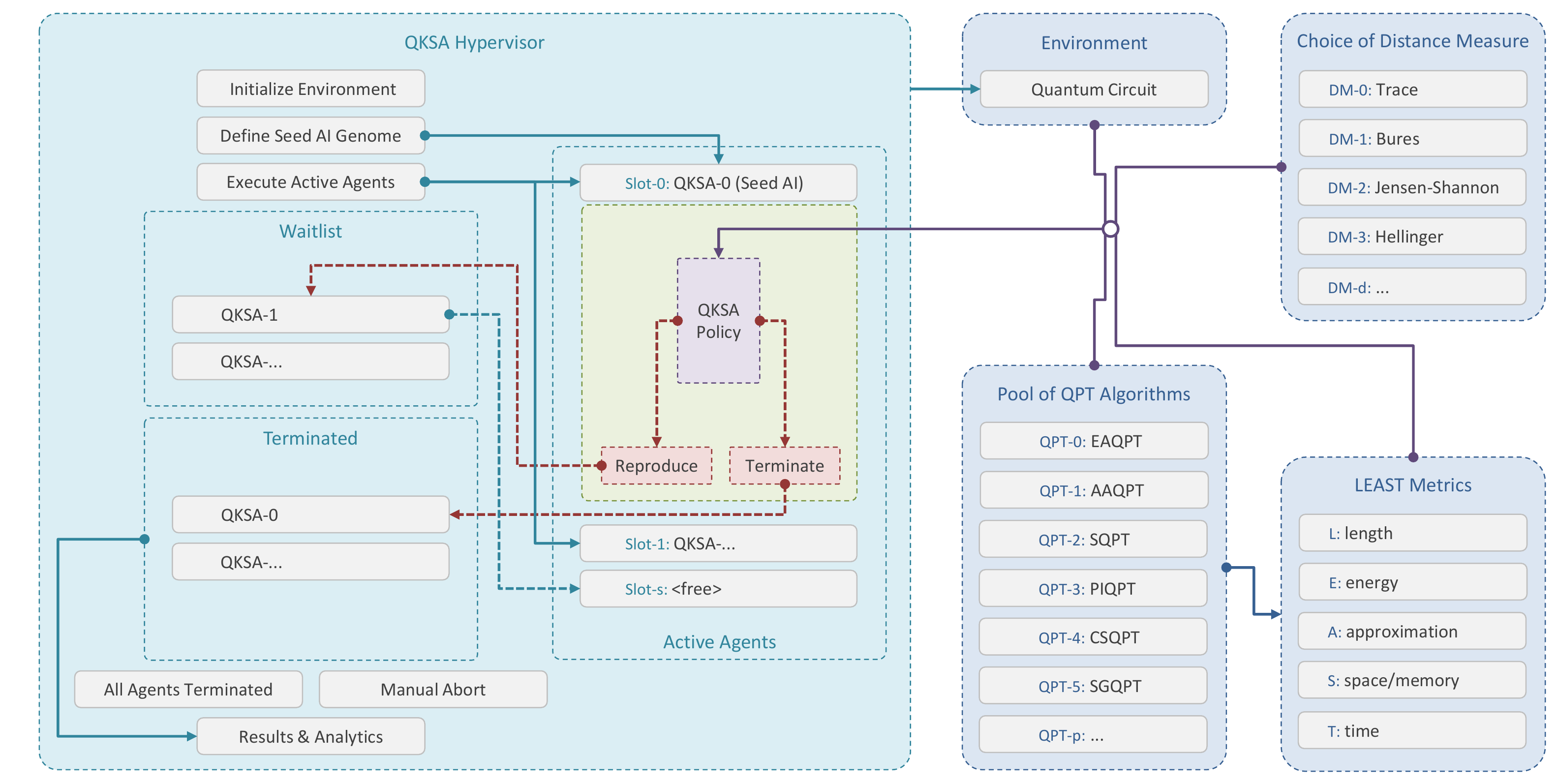}
\caption{QKSA framework}
\label{fig:modules_l1}
\end{figure}

The QKSA framework as shown in Figure~\ref{fig:modules_l1}, consists of 5 major blocks: environment, pool of QPT algorithms, LEAST metrics cost estimators, choice of distance measure and the QKSA hypervisor.
The execution procedure and the interaction between these blocks are explained in this section.

The environment is defined by the user as a Qiskit quantum circuit.
The QKSA also allows probabilistic mixtures of quantum circuits and partially observable environments.
Currently, only episodic environments are considered.
Thus, each cycle of agent interaction resets the environment based on the focus of QPT.
In non-episodic environments, the Holevo bound restricts the total classical bits of information that can be extracted, limiting the applicability in model-based reinforcement learning.
The environment is shared between the agents.
Each agent can choose to measure only a part of the shared environment, thus can be used for studying non-local strategies.
The environment also defines the set of actions $\mathcal{A}$ and perceptions $\mathcal{E}$ that can be used by the agent for interaction.
This set is defined automatically from the number of qubits used to define the environment.
The user can however modify and restrict the set based on the intended purpose, e.g.\ only Z-basis measurements are required for studying quantum versions of classical logic like a quantum adder.

The second block consists of a pool of QPT algorithms.
Each algorithm is capable of taking as input the history $h_t$ and output the environment model $\rho_t$.
Any new QPT strategy can be coded and added in the pool as a black-box algorithm as long as this criteria is met.
Note that initially the history is an empty sequence.
This corresponds to a maximally mixed density matrix.

Each QPT algorithm can also be evaluated for the LEAST metrics based on the cost estimators.
The cost estimators use both online and offline methods to estimate the cost.
For example, the length and approximation estimate of the QPT algorithm can be directly inferred from inspecting the program code, while the run-time is estimated while the optimal action is being evaluated.

The framework also offers a pool of distance metrics $\Delta$ between quantum density matrices $\rho$.
The goal of QKSA is different from QPT for device characterization as the environment is unknown.
Thus, the distance between the current model and the actual model cannot be calculated.
Instead, the metric measure between the predicted model update and the actual update is used to infer the learning gradient.
From our experiments, we provide a set of metrics that have a monotone behavior and is a distance measure for quantum processes.
However, there are many distance measures and the user can choose a specific default measure or let each agent randomly choose one during the initialization.

The last module is the hypervisor that encapsulates the QKSAs.
The seed QKSA constitutes the minimal implementation of the QKSA.
This agent is instantiated by the QKSA hypervisor and added to the active pool of agents.
Thereafter, the hypervisor executes each active agent, either in parallel or by dovetailing.
Each agent learns the environment based on its own policy.
When the learning converges, the agents reproduce by mutating their policy.
The new agents are added to the waitlist and are automatically instantiated by the hypervisor when computational resources are available.
Eventually the agents completely learn the environment (or the maximum lifetime limit is reached).
Then the agent is terminated.
The user can also manually terminate all active agents.
Thereafter, the learning results of each active/terminated agent are displayed for analysis.

\begin{figure}[h]
\centering
\includegraphics[clip, trim=47cm 0cm 0cm 53cm,width=0.99\textwidth]{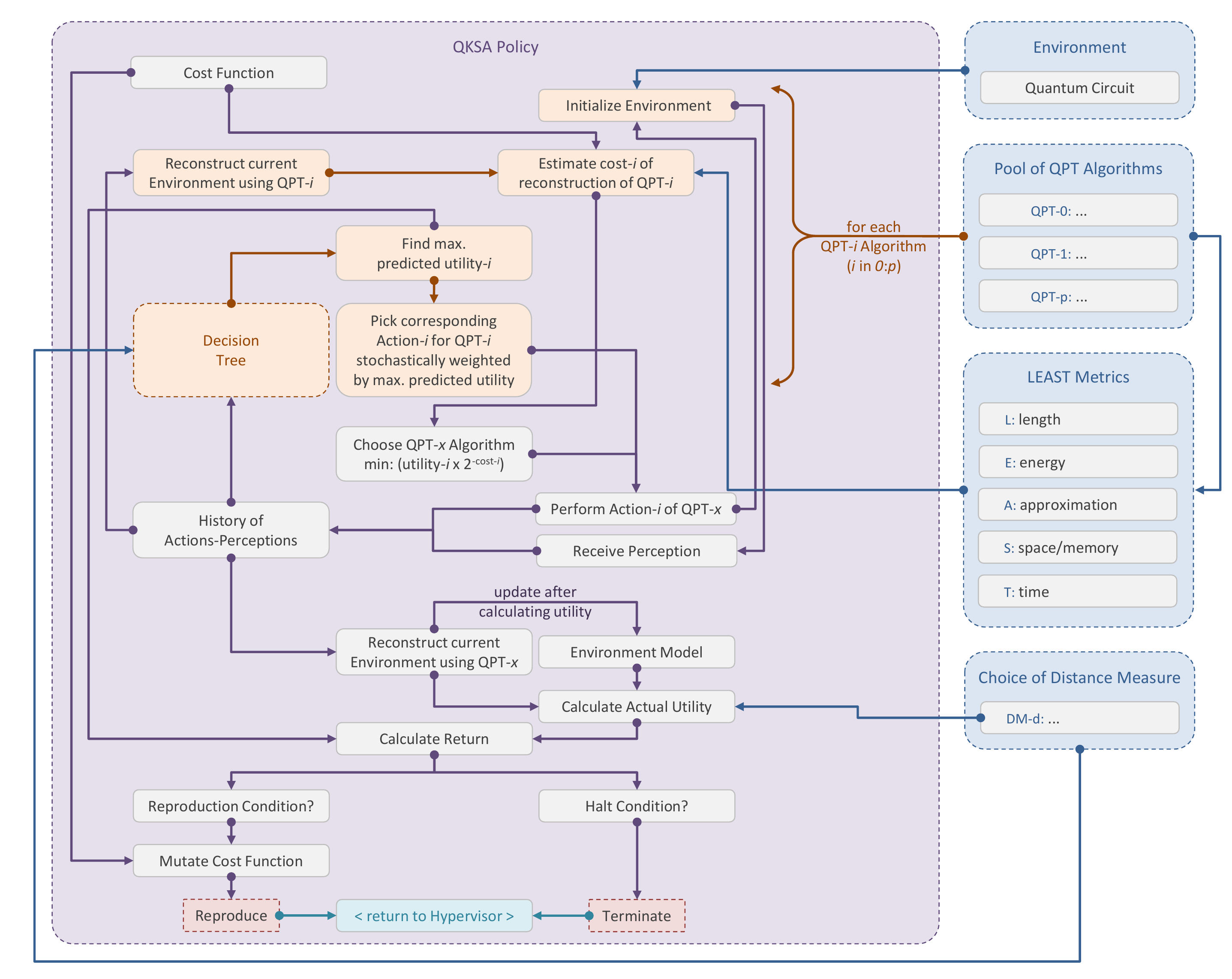}
\caption{QKSA policy}
\label{fig:modules_l2}
\end{figure}

Now let us describe the QKSA policy in further depth.
This is depicted in Figure~\ref{fig:modules_l2}.

Each QKSA has its own Cost Function, which is its part of the mutable genome.
Other parameters are passed to the QKSA from the hypervisor (for the seed QKSA) or the parent via immutable genes in the remaining part of the genome.
These parameters are used to initialize the agent.
At this stage, the quantum circuit for each available QPT algorithms is also created.
This might involve additional initialization circuits for the QPT strategy (e.g. entangling ancilla for EAQPT).

Then for each QPT algorithm, the best action is evaluated based on the corresponding predicted utility.
This utility is based on the distance between the current environment model and the predicted environment model based on the chosen action and probabilistic perception.
The LEAST cost for the QPT is also evaluated while the reconstruction is done.
After this is done for all QPT algorithms, the weighted (by the cost) maximum utility is used to pick a QPT strategy.

The action of the corresponding chosen QPT is performed and the perception is received from the modified environment of the specific QPT algorithm.
This is used to calculate the actual utility as well as update the current model of the environment and the history.

The difference between the predicted and the actual utility is the return.
This value is used to determine the learning progress and trigger the reproduction or termination of the agent.

The learning routine first evaluates the cost of each QPT strategy based on the cost estimators for choosing the current optimal action.
QPT strategies that are beyond the allowed threshold are filtered out.
A weighted selection is done for a specific QPT algorithm based on the cost of the remaining strategies.
This defines the actual action to the environment and the prediction from the chosen QPT algorithm.
The actual perception from the environment is used to update the model and calculate the utility for the current prediction based on the distance metric for the agent.
The utility over the past steps is evaluated to assess the return.
When the utility falls below a threshold, the return is used to determine the fitness of the agent.
If the agent is fit, it replicates with mutation on the cost function, otherwise it halts.
In the standard default setup, the only difference between agents is the cost function.
The replication is carried out by invoking the mutating quine subroutine.
The new program file for the child QKSA is automatically instantiated by the hypervisor.
The parent quine after reproduction continues predicting the environment and reproducing.
An upper bound on the number of replications is set after which the agent is archived by the hypervisor.

\section{Experimental results} \label{s5}

A full proof-of-concept of the discussed QKSA framework is implemented on Python and Qiskit.
It is available as an open-sourced software on GitHub at the following link: https://github.com/Advanced-Research-Centre/QKSA.
In this section we present an initial experiment that demonstrates the features of QKSA as presented in the previous sections.

In this experiment we consider the choice between two QPT strategies, both EAQPT.
QPT-0 has an approximation of 5 decimal places while using 16384 steps of history.
QPT-1 has an approximation of 8 decimal places however uses only 8192 steps.
Since they are the same algorithm, the program length variance is negligible.
The time to reconstruct for QPT-0 is more because of the history (though the coarser approximation reduces it a little).
Given such a description, it is not immediately clear which QPT would work best to reconstruct and model a given quantum environment.
It depends also on how simple/complex the environment is (e.g. if the additional decimal places have useful information).
In such situations, QKSA can be readily applied.

\begin{figure}[h]
\centering
\includegraphics[clip, trim=7cm 0cm 0cm 3cm,width=0.99\textwidth]{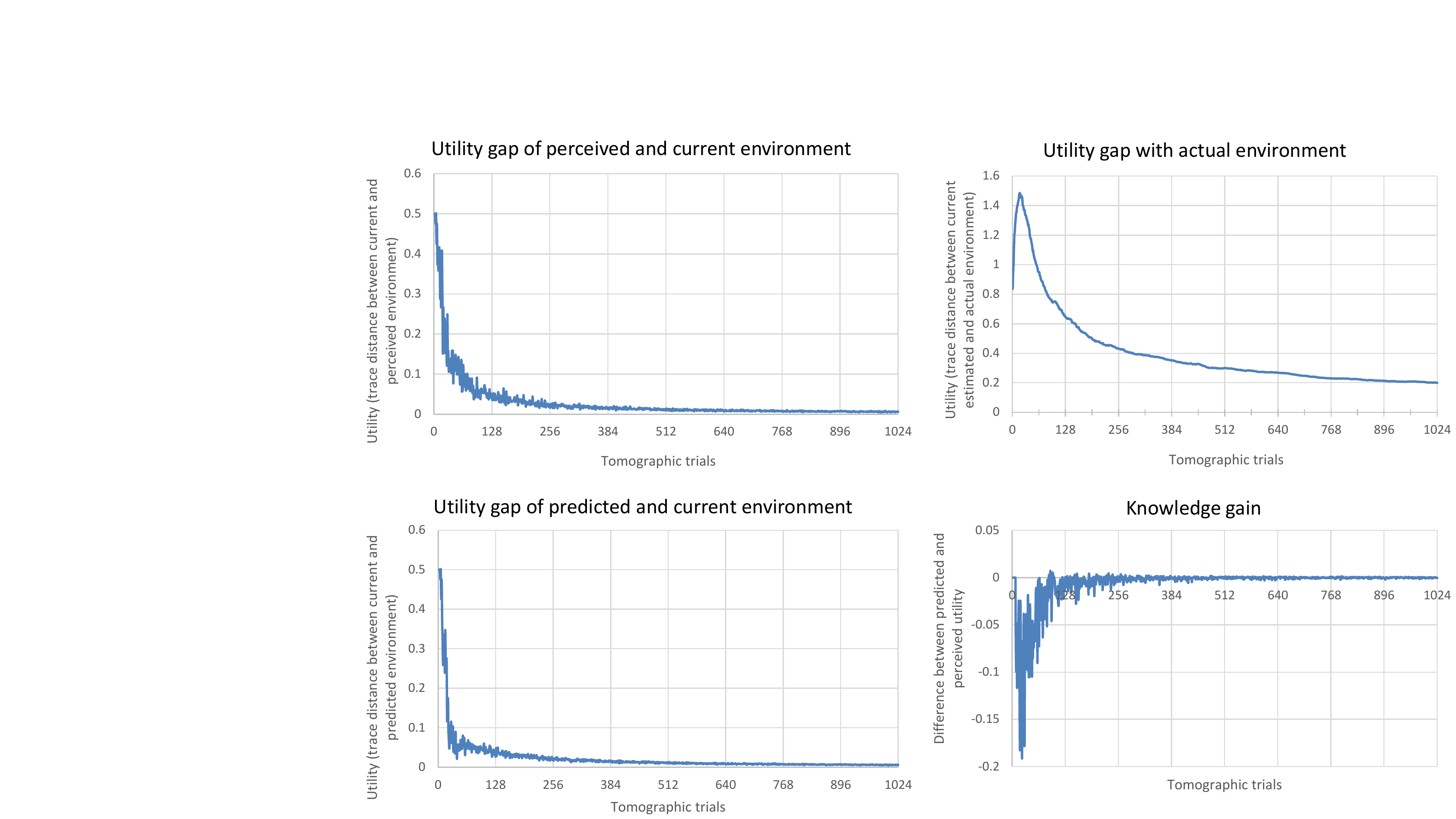}
\caption{Trace distance EAQPT 1-qubit environment random unitary average of 20 experiments}
\label{fig:res2}
\end{figure}

We defined a random 1-qubit unitary as the environment.
Results of averaging over 20 random circuits are shown in Figure \ref{fig:res2}.
The chosen distance measure chosen is trace distance.
We found that Bures distance and Hilbert-Schmidt distance do not perform well.
This is because the distance between the initial complete mixed state and the Choi matrix for the unitary in EAQPT is close to zero.
The perceived and predicted utility are plotted on top-left and bottom-left respectively.
The perceived utility is the actual information gained by the agent on performing the action-perception interaction for the step.
The predicted utility is the quantum generalization of the utility of KL-KSA.
In this context, the quantum process represented as the Choi density matrix is the compressed representation of the environment.
The difference between the predicted and perceived utility is the knowledge which reflect how well the current model agrees with the actual environment.
In the top-right subplot we show the remaining utility with the time step.
It is only possible to know this when the target environment is known.
While this is not the case for QKSA, we plot this to show the convergence of the learning behavior.


We show in the results that the perceived utility converges to zero.
The striking advantage of this approach is that, without knowing the actual environment, this trend can be used to optimize a quantum algorithm.
This can have significant impact in algorithms like VQE where quantum tomography is an integral part.
Current tomography routines set a constant value of trial runs on the quantum computer which is significantly more than the needed trials for typical statistical approximation threshold requirements.
Via this online evaluation of the environment, it is possible to predict when the environment has been sufficiently modeled and abort the learning process.
This can also be useful when the noise characteristic is temporally variant during the execution of the algorithm.

\section{Conclusion} \label{s6}

In this work we extended the formalism of universal artificial general intelligence (UAGI) to quantum environments.
We generalized the KL-KSA to a quantum knowledge seeking agent (QKSA).
The environment within the reinforcement learning setup is defined by an unknown quantum circuit in Qiskit.
The agent models the environment using quantum process tomography algorithms.
A quantum environment prevents the exact prediction of perceptions (as used by AIXI), as well as a single probability distribution of perception based on the set of actions (as used by KL-KSA).
The probability distribution is conditioned on the chosen action, and is thus represented by the more general density matrix formalism.
Any quantum process can be represented as a Choi density matrix, which forms a model of the environmental dynamics.

Despite their theoretical significant, UAGI models are uncomputable, thus are not useful for practical learning tasks.
A typical solution is to restrict the runtime and length of the programs.
Such solutions have been shown to learn simple games like Pac-Man.
However, the space of programs grows exponentially, and thus a simple cutoff is not a scalable solution.
To circumvent this, we propose to evaluate the algorithmic cost within a set of user provided Python codes instead of enumerating Turing machines.
This considerably makes the framework more tractable.

Finally, the resource restrictions used in computable UAGI models (like UCAI and AIXI-tl) are arbitrary.
In our model, these resource bounds are interdependent hyper-parameters whose value and trade-off relations are optimized using genetic programming.
Thus, this allows open-ended evolution of the agents for changing environments.
Each agent can self-replicate as a quine and thus is a recursive self-improving model of intelligence.

QKSA provides a framework to evaluate a swarm of UAGI agents that discover the resource tradeoffs on modeling a quantum environment.
Besides the theoretical importance, the QKSA framework can be used to study the applicability of various distance measures of quantum information.
It also has near term applicability in optimizing NISQ era variational quantum algorithms like QAOA which rely on multiple runs of quantum tomography.
We show as a proof-of-concept that it can be used in quantum process tomography where the QKSA knowledge gain reflects the trace distance with the unknown environment.

As part of currently ongoing research, we are applying the QKSA framework as described in this article to study course-graining in multiple observers scenarios and quantum uncomplexity resources.



\let\oldbibliography\thebibliography
\renewcommand{\thebibliography}[1]{%
  \oldbibliography{#1}%
  \setlength{\itemsep}{0.8pt}%
}

\bibliographystyle{unsrt}
\bibliography{ref2.bib}

\end{document}